\documentclass[intlimits,twoside,a4paper]{article}

\usepackage{amsmath,amssymb}
\usepackage{graphicx}

\usepackage[T2A]{fontenc}
\usepackage[cp1251]{inputenc}

\usepackage[eqsecnum]{cmpj2}

\issue{2016}{19}{4}{43603}
\doinumber{10.5488/CMP.19.43603}
\title[Grain boundary structure and local magnetism]%
{Grain boundary relaxation and reconstruction: effect on local magnetic moment}

  \author[E. Vitkovsk\'{a}, P. Ballo]{E. Vitkovsk\'{a}\thanks{E-mail: atavaha@hotmail.com}\,, P. Ballo}
  \address{Slovak University of Technology, 3 Ilkovi\v{c}ova, 812 19 Bratislava, Slovakia}

\authorcopyright{E. Vitkovsk\'{a}, P. Ballo, 2016}
\date{Received August 4, 2016}

\begin{document}

\maketitle

\begin{abstract}
We present a detailed numerical study on structure and local magnetic properties of $\langle 100 \rangle$ symmetric tilt grain boundaries in bcc-iron. Particular attention is paid to connection between type of grain boundary relaxation and local magnetic properties. Results from first principles calculation showed that grain boundary reconstruction leads to non-uniform distribution of local magnetic moments in grain boundary plane. This is in contrast with the result obtained in grain boundary plane, where simple relaxation is observed. Well optimized atomic configurations in the vicinity of the interface were achieved by  simulated annealing optimization technique improved by combination with genetic algorithm.
\pacs 61.72.Mm, 75.50.Bb, 31.15.E-, 02.70.Tt
\end{abstract}

\section{Introduction}
\label{S:1}
Grain boundaries (GBs) as interfaces between two grains significantly influence wide range of physical properties of polycrystalline and nano-crystalline materials. From microscopical point of view, the distorted crystal structure along the GBs is mainly responsible for changes in various physical properties of materials \cite{mishin10}. In case of ferromagnetic materials, there is also connection between the distorted structure and local magnetic properties \cite{siegel05,wachowicz08}. Recently, Ii et al. \cite{ii13} for the first time made direct measurements of local magnetic moments at grain boundaries in iron. They used electron energy loss spectroscopy on a transmission electron microscope. Since then there have been several attempts to simulate local magnetic moment at iron grain boundaries from first principles \cite{hampel93,wu96,cak08,wachowicz10}. The increase of local magnetic moment on GB plane by 15 to 18\% due to magneto-volume effect and oscillatory behaviour in planes parallel to GB explained by Stoner model \cite{stoner38,manh09} was reported. The results were achieved on symmetric tilt GBs: $\Sigma 5(310)$ \cite{hampel93,cak08}, $\Sigma 3(111)$ \cite{wu96,wachowicz10} and $\Sigma 5(210)$ \cite{wachowicz10}. The experimental study by Ii et al. \cite{ii13} confirmed the increase of local magnetic moment at grain boundaries. Moreover, the local magnetic moment showed an increasing trend up to misorientation angle $45^{\circ}$ and  the cusps on the magnetic moment, i.e., misorientation angle curve at grain boundaries with low coincidence site density, specifically $\Sigma 9$.

The inevitable part with non-negligible effect on the result is GB type and structure. Consequently, structure optimization and detailed description of optimized GB structure is an essential part of the study of GB properties. Optimization process invokes either relaxation or reconstruction of the ideal geometrical boundary. Relaxation and reconstruction are terms especially connected with free surfaces \cite{stekol02,crljen03}. However, the use of these terms in case of GBs is also reasonable \cite{ras12,persson03,yang15,morris96}. The most common relaxation observed is oscillatory behaviour of inter-planar distances in the direction perpendicular to GB \cite{du11,wachowicz10,cak08}. The oscillatory behaviour can be combined with small atomic displacements within planes parallel to GB plane. Whenever the structure of grain boundary plane differs from the structure of the corresponding lattice planes in the grain, one speaks of a reconstructed GB structure. Reconstruction is a result of complicated series of atomic displacements leading to the change of period or coordination number \cite{morris96}, migration of GB plane \cite{yang15}, formation of vacancies or anti-site defects \cite{ras12}, etc.

Due to the high computational demands, first principles calculations use more simple gradient optimization methods. While the relaxation is easily achieved by ``simple'' optimization techniques, the process seems to be much more complicated in the case of reconstruction. Recently, the application of genetic algorithms (GA) to search stable atomic structures \cite{zhang09,dugan09,amy15} has identified GA as very effective structure optimization technique.

The aim of our work is to investigate local magnetic behaviour of four bcc-iron $\langle100\rangle$ symmetric tilt GBs: $\Sigma 5(210)$, $\Sigma 5(310)$, $\Sigma 17(410)$ and $\Sigma 13(510)$. The first two as representatives with a high coincidence site density and the last two as representatives with a low coincidence site density. The investigated GBs cover the range of misorientation angle from $22.6^{\circ}$ to $53.1^{\circ}$. To ensure well-optimized structures of the investigated GBs, we improved well-known simulated annealing (SA) \cite{kirk83} technique by addition of some GA features. Inter-atomic interaction is in the process of optimization described by Embedded-atom method (EAM) \cite{daw84} and final optimized structures act as an input to first principles calculation. The article is arranged as follows: GB supercell construction, new optimization method and first principles simulation details are described in section \ref{S:2}. In section \ref{S:3} the main results are presented and discussed. Finally, the main contribution of our work is summarized in section \ref{S:4}.
\section{Computational method}
\label{S:2}
\subsection{Grain boundary model}
\label{S:2.1}

\begin{figure}[!b]
\centerline{\includegraphics[width=0.65\textwidth]{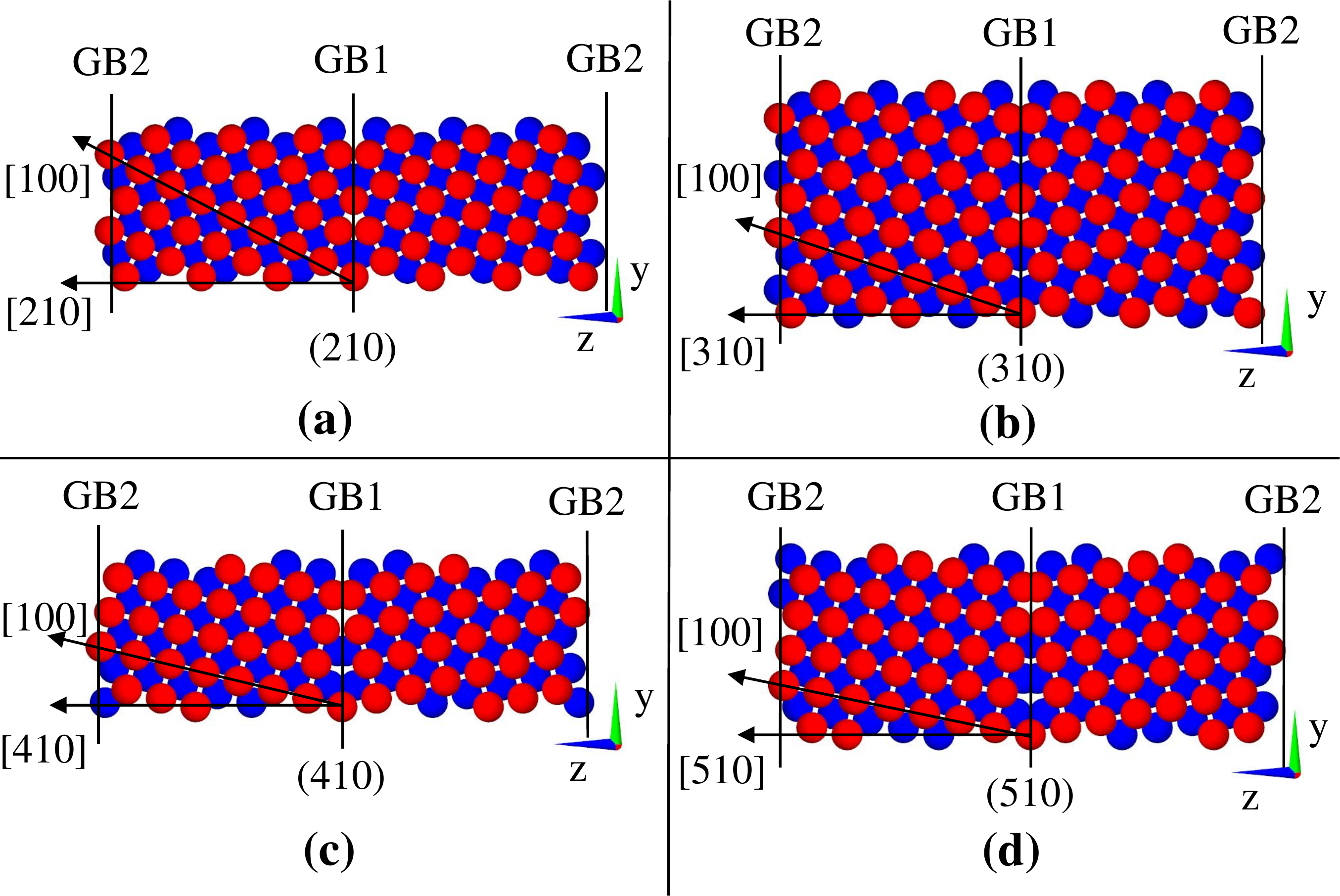}}
 \caption{(Color online) Models of bcc-iron $\langle100\rangle$ symmetric tilt grain boundaries (a) $\Sigma 5(210)$, (b) $\Sigma 5(310)$, (c) $\Sigma 17(410)$, (d) $\Sigma 13(510)$. Red and blue colour represents atoms placed in planes $(001)$ and $(002)$, respectively.}\label{fig1}
\end{figure}

	Grain boundary supercells were constructed by rotating the two grains in the opposite direction by the same angle around the $[001]$ rotation axis and then matching the two grains together. The result of this is that in the direction of boundary plane the GB shows periodic structure. Atomic positions in the computational supercell are generated according to coincidence site lattice theory. This geometry in combination with periodical boundary conditions creates  computational cell  which  contains two GBs. The first of them is positioned in the middle of the computational cell (GB1) while the second  (GB2) is  from geometrical reasons split into two parts positioned on the top and bottom edge of the cell. Model of computational supercells is shown in figure~\ref{fig1}. GB structure was optimized by our improved optimization technique, which we will refer to as OPSA (optimization by parallel simulated annealing), in combination with EAM. First principles calculation was applied only for magnetic moment computation. Note, two different types of supercells were used for GB structure analysis and for magnetic moment calculations. Due to high computational demands of first principles simulations, the dimensions of supercells designed for magnetic calculation were cropped to provide maximum 100~atoms per supercell. Supercell details are summarized in table~\ref{tab1}. Original and cropped GBs were optimized separately, so in both cases both grain boundaries GB1 and GB2 were optimized and, therefore, are equivalent.
\begin{table}[!h]
\caption{Parameters of bcc-iron symmetric tilt $\langle 100 \rangle$ grain  boundary simulation supercells used for structural analysis (EAM) and magnetic moment calculation (\textit{ab initio}): misorientation angle ($\alpha$),  dimensions in $x$, $y$, $z$ direction ($sx$, $sy$, $sz$), number of atoms ($N$). } \label{tab1}
\vspace{2ex}
\begin{center}
\renewcommand{\arraystretch}{0}
\begin{tabular}{|l|c|c|c|c|c|c|c|c|c|}
\hline\hline
\multicolumn{2}{|c|}{Type} & \multicolumn{4}{c|}{EAM} &\multicolumn{4}{c|}{\textit{ab initio}} \strut\\
\hline
GB & $\alpha[^{\circ}]$ & $sx$(\AA)& $sy$(\AA)& $sz$(\AA)& $N$  &
                          $sx$(\AA)& $sy$(\AA)& $sz$(\AA)& $N$  \strut\\
\hline
\hline
$\Sigma 5(210)$ & 53.13 & 14.28 & 12.77 & 40.86 & 640  &
                           2.86 &  6.38 & 40.86 & 64   \strut\\
                           \hline
$\Sigma 5(310)$ & 36.87 & 17.13 & 18.07 & 37.92 & 1008 &
                           2.86 &  9.03 & 37.92 & 84   \strut\\
                           \hline
$\Sigma 17(410)$& 28.07 & 11.42 & 11.77 & 38.78 & 448  &
                           2.86 & 11.77 & 33.24 & 96   \strut\\
                           \hline
$\Sigma 13(510)$& 22.62 & 14.28 & 14.56 & 39.19 & 700  &
                           2.86 & 14.56 & 27.00 & 100  \strut\\
\hline\hline
\end{tabular}
\renewcommand{\arraystretch}{1}
\end{center}\vspace{-2mm}
\end{table}

\subsection{Optimization technique}	
	 Structures of GBs were optimized by our newly improved and parallelized SA optimization algorithm OPSA.  The basic strategy of optimization is based on the technique of SA. It was proved that by carefully controlling the rate of cooling the temperature, SA can find the global optimum. However, this requires infinitely long computational time. In real simulation, the result strongly depends on experimental details, like random generator seed. In the case of GBs, which are in discussion, there are many metastable configurations separated by relatively high energetic barriers. The existence of barriers makes the problem complicated and it is almost impossible to solve it in real time.
	
To avoid this problem, we applied the technique of parallelization which is based on the idea of many simultaneous runs of identical problem on parallel CPUs. Parallelization was included via certain GA features. OPSA works with a group of individuals (supercells with different GB structure) which are individually annealed in several generations (annealing cascades) ending with crossover operation. The whole optimization process was carried out in detail as follows:

\looseness=-1 At the first step, we generate the geometric position of atoms positioned in the computing cell. The process by which the positions of atoms were generated is described in section~\ref{S:2.1}. The data are multiplied $N$-times where $N$ is the number of processors on which the parallel job is running. Subsequently, atoms positioned in the vicinity of GBs are slightly shifted using the random numbers generator so that displacement of individual atom does not exceed $0.5$~\AA ~in each direction. Next step is SA which runs parallel on CPUs for different individuals. The process runs from $350$~K temperature to $7.6$~K using a stepwise exponential decrease of temperature involving a total of $250$ steps. At the end, temperature of $0$~K is reached by an acceptance of the position of atoms with lower energy. This step corresponds to zero generation.

At the second step, we create a next generation. The next generation is created by the SA process which runs from temperature $100$~K to temperature $7.7$~K in $100$ steps. The process is finalized with $0$~K temperature like in the previous step. The result is a new generation of $N$ individuals.

The third step was motivated from the experiences stating that in the case of complex interface, a group of atoms can occur into local minima that cannot be overcome by the method of SA. The group of these atoms increases the overall energy of GB. To solve this problem, we proceed as follows: after the end of SA in the second step, the energy of each GB is determined and  compared with the previous GB energy. If the energy of GB after annealing does not reduce by more than a threshold value (in our case it is $0.005$~J\,m$^{-2}$), this individual is labelled as stuck, and the group of atoms that has caused it is identified as the critical group. Subsequently there is selected the same group of atoms consisting of a critical group located in a successful individual. Successful individual is the individual, whose critical group of atoms has the lowest overall energy. This group of atoms replaces the stuck group of atoms in an unsuccessful individual. Identification is based on the position of the central atom of the critical group of atoms. This is the procedure of mating which is a part of GA. Having carried out this, the algorithm loops back to the second step and repeats this procedure until the desired result is reached.

The energy of GB, which is a very important parameter for the assessment of the optimization, is defined as follows:
\begin{equation}
\label{eq:egb}
 E_\text{{GB}}=\frac{E_\text{{GB}}-NE_\text{{C}}}{2s_{x}s_{y}}\,,
\end{equation}
where $E_\text{{GB}}$ is the energy of a supercell with two optimized GBs, $N$ is the number of atoms in a supercell, $E_\text{{C}}$ is the cohesive energy and $s_{x}s_{y}$ is GB area. The energies were calculated using EAM potential with  parametrization provided by Mendelev et al. \cite{mendelev03} According to this, parametrization was a lattice parameter for bcc-iron set up to $2.855$~\AA.

	The described method was tested on four GBs, which are considered in our study. Details about dimensions, geometry and the number of atoms in supercells which were used in this calculation are summarized in table~\ref{tab1}. We refer to these supercells as SMALL.  As a comparison, we performed the same simulation on  extended supercells with four times more atoms in GB plane. We will refer to these supercells as LARGE. The energies of GBs obtained by standard SA and OPSA algorithm are summarized in table~\ref{tab2}. Note that standard SA corresponds to the first step in the OPSA algorithm. The number of individuals was set to 15 for SMALL and to $23$ for LARGE supercell. The number of generations was limited to $50$. The results show that OPSA algorithm yields better results for all cases compared to SA algorithm. However, in the case of LARGE supercell, the efficiency of OPSA algorithm increases. This can be explained by a higher incidence of local minima which incurred as a result of the extension of GB. We must emphasize that the result for SA is the best one out of $15$ (SMALL), $23$ (LARGE) individuals. The results in table~\ref{tab2} also show that while relaxation is easily achieved by standard optimization techniques like SA, reconstruction is not. LARGE GBs $\Sigma 17(410)$ and $\Sigma 13(510)$ remained after SA optimization in an unreconstructed form with unacceptably high energy around $2$~J\,m$^{-2}$. The reconstruction was not fully accomplished also for SMALL GB $\Sigma 13(510)$. Therefore, the optimization technique is an essential part of GB study with a non-negligible impact on the result.
\begin{table}[!h]
\caption{Comparison of grain boundary (GB) energies (in J\,m$^{-2}$) obtained by SA and OPSA algorithms. GBs referred to as LARGE have 4-times larger GB area than SMALL ones. Significant values are bold-type.}\label{tab2}
\vspace{2ex}
\begin{center}
\renewcommand{\arraystretch}{0}
 \begin{tabular}{|l||c|c|c|c|}
 \hline\hline
 \multicolumn{1}{|c||}{GB} &\multicolumn{2}{c|}{SMALL} &\multicolumn{2}{c|}{LARGE} \strut\\
 \hline
 Optimization    &SA 350~K & OPSA & SA 350~K & OPSA  \strut\\
 \hline
 \hline
 $\Sigma 5(210) $& 1.462  & 1.462 & 1.600 &  1.469\strut\\
\hline\
 $\Sigma 5(310) $& 1.054  & 1.053 & 1.060 & 1.059  \strut\\
\hline
 $\Sigma 17(410)$& 1.235  & 1.231 & \textbf{1.969} & \textbf{1.280}  \strut\\
\hline
 $\Sigma 13(510)$& \textbf{1.402}  & \textbf{1.070} & \textbf{2.022} & \textbf{1.191}  \strut\\
\hline\hline
 \end{tabular}
\renewcommand{\arraystretch}{1}
\end{center}
\end{table}
\subsection{First principles calculations}
\begin{figure}[!b]
\centerline{\includegraphics[width=0.68\textwidth]{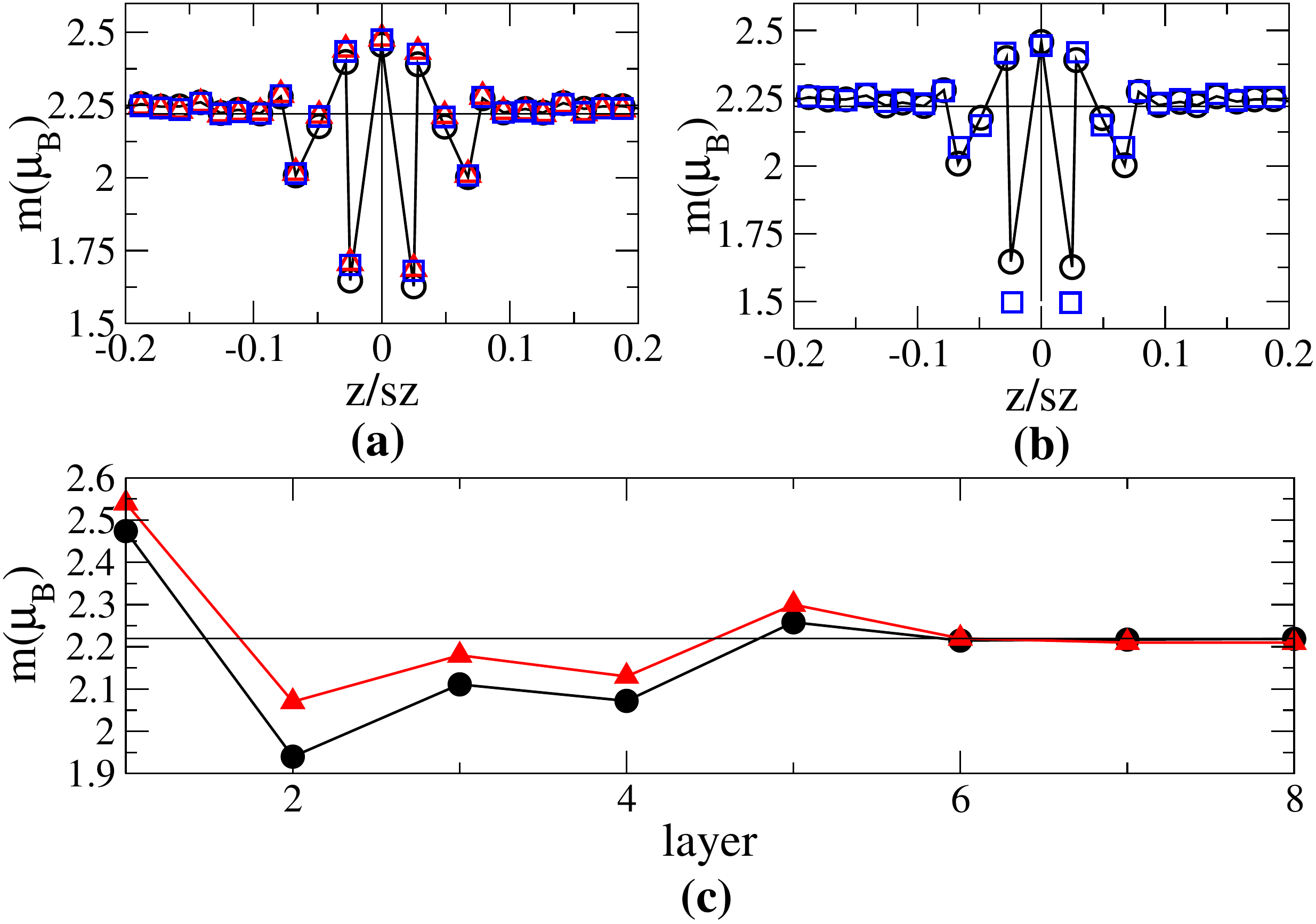}}
 \caption{(Color online) Variation of magnetic moment on Fe atoms in the direction perpendicular to the GB $\Sigma 5(210)$ (a) computed with 4~$k$-points (circles), 8~$k$-points (triangles), 16~$k$-points (squares) (b) computed withouth (circles) and with (squares) additional first principles optimization. (c) Local magnetic moments of Fe atoms in the neighbourhood of GB $\Sigma 5(310)$, our result (circles) and result obtained by \v{C}\'{a}k~et~al. \cite{cak08} (triangles). }\label{fig-a1}
\end{figure}

	First principles calculations were carried out using the density functional theory (DFT) formalism as implemented in the ABINIT code \cite{gonze09}. The electron-ion interaction was described by the norm-conserving Troullier-Martins pseudopotential \cite{troullier91}. The exchange correlation energy was treated in the local density approximation (LDA) using Teter-Pade parametrization \cite{goedecker96}. Before magnetic moment computation, a series of convergence tests were performed and basic bcc-iron parameters were computed. Results are listed in table~\ref{tab3}. For comparison, results obtained by our EAM simulation and experimental values are listed in table~\ref{tab3}.  $K$-point space was sampled by $2\times2\times2$ Monkhorst-Pack set, which corresponds to 4~$k$-points. The effect of $k$-point sampling on the quality of local magnetic moment was intensively tested on GB $\Sigma5(210)$. Results which are shown in figure~\ref{fig-a1}~(a) demonstrate that the number of 4~$k$-points used provides a sufficient quality and does not need to be increased. The cut-off energy of the plane wave set was set to $1632$~eV. Thermal broadening was defined as Fermi-Dirac smearing with temperature of $0.27$~eV. It should be noted that the lattice parameter was changed from EAM equilibrium value $2.855$~\AA~ to DFT equilibrium value $2.823$~\AA~ and no additional relaxation was applied within first principles calculation. The effect of additional first principles optimization was tested on GB $\Sigma5(210)$, as the most unstable one, concerning the highest GB energy connected with this GB. The comparison of local magnetic moment behaviour without and with additional BFGS quasi-Newton method optimization performed within first principles calculation is in figure~\ref{fig-a1}~(b). The result in figure~\ref{fig-a1}~(b) shows a very good coincidence in the qualitative magnetic moment behaviour and only a small (maximum 6\% related to bulk value) quantitative change. Another argument supporting good performance of combination of EAM~+~OPSA optimization and first principles magnetic moment calculation is a very good agreement between our calculations and local magnetic behaviour on GB $\Sigma5(310)$ presented by \v{C}\'{a}k et al. \cite{cak08}. Comparison is in figure~\ref{fig-a1}~(c). The advantage of our approach lies in a very good performance to computational demand ratio. While first principles optimization takes several days, OPSA~+~EAM optimization can be achieved within one day.
\begin{table}[!t]
\vspace{-1mm}
\caption{Basic bcc-iron parameters computed by EAM and DFT calculation compared with experimental values: lattice parameter ($a$), cohesive energy ($E_\text{{C}}$), bulk modulus ($B$), magnetic moment ($m$), magnetic moment of free iron atom ($m_\text{{free}}$).} \label{tab3}
\vspace{2ex}
\begin{center}
\renewcommand{\arraystretch}{0}
 \begin{tabular}{|l||c|c|c|}
\hline \hline
 Parameter       & EAM & DFT & Experiment \cite{kittel96,pearson67} \strut\\
 \hline
 \hline
 $a$[\AA]           & 2.855 & 2.823 & 2.286 \strut\\
 \hline
 $E_\textrm{{C}}$[eV]        &$-$4.122 &$-$5.14  &$-$4.28   \strut\\
 \hline
 $B$[GPa]           & 178   & 201   & 180    \strut\\
 \hline
 $m$[$\mu_\textrm{B}$]       & $-$     & 2.1   & 2.22   \strut\\
 \hline
 $m$$_\textrm{{free}}$[$\mu_\textrm{B}$]& $-$     & 3.97  & 4.0    \strut\\
 \hline\hline
 \end{tabular}
\renewcommand{\arraystretch}{1}
\end{center}
\vspace{-3mm}
\end{table}

$K$-point sampling has a significant impact on the results obtained by DFT calculation. In order to compute GB energy by DFT, we prepared a special supercell. The supercell contains 100 atoms arranged into ideal bcc-crystal. The dimensions of the supercell are $2.823~\AA\times14.1150~\AA\times28.23~\AA$, which ensures the shape consistent with GB-supercells described in section~\ref{S:2.1}. $K$-point mesh for a special supercell was adopted from simulation with GB-supercell. In this instance, GB energy is computed as follows:
\begin{equation}
\label{eq:egb2}
 E_\text{{GB}}=\frac{E_\text{{GB}}-\frac{1}{100}NE_{100}}{2s_{x}s_{y}}\,,
\end{equation}
where $E_\text{{GB}}$ is energy of supercell containing two GBs, $E_{100}$ is energy of special 100-atom supercell, $N$ is the number of atoms within GB-supercell and $s_{x}s_{y}$ is GB area. Comparison of GB energies obtained by EAM and DFT simulations is in table~\ref{tab4}. In comparison with EAM, DFT values are systematically overestimated up to $0.8$~J\,m$^{-2}$.
\begin{table}[!h]
\begin{center}
\renewcommand{\arraystretch}{0}
\caption{Comparison of grain boundary (GB) energies (in J\,m$^{-2}$) obtained by EAM and DFT simulation. $\Delta$~indicates GB energies related to GB energy of GB $\Sigma 5(310)$. } \label{tab4}
\vspace{2ex}
 \begin{tabular}{|l||c|c|c|c|}
 \hline\hline
 GB               & EAM   & DFT   & $\Delta$EAM & $\Delta$DFT \strut\\ \hline
 \hline
 $\Sigma  5(210)$ & 1.468 & 2.273 &   $+$0.418    & $+$0.548 \strut\\
 \hline
 $\Sigma  5(310)$ & 1.050 & 1.725 &   $-$         & $-$      \strut\\
 \hline
 $\Sigma 17(410)$ & 1.131 & 1.796 &   $+$0.081    & $+$0.071 \strut\\
 \hline
 $\Sigma 13(510)$ & 1.124 & 1.752 &   $+$0.074    & $+$0.027 \strut\\
 \hline\hline
 \end{tabular}
\renewcommand{\arraystretch}{1}
\end{center}
\end{table}
\vspace{-3mm}
 \section{Results and discussion}
\label{S:3}

	The energy of individual GBs in the process of optimization can be reduced in several ways. In particular, by the change of inter-planar distances between planes parallel to GB, by a rigid
shift of one grain with respect to another, or by small atomic displacements within planes parallel to GB. In this case, we define the result of optimization process as relaxation, where various atomic positions are separated only by a low energetic barrier and can be described in terms of the changed inter-atomic distances or bond angles. The result is mostly independent of the choice of the optimization method. The relaxation mechanism was observed in the case of GBs $\Sigma 5(210)$ and $\Sigma 5(310)$. In both cases there was identified a shift of inter-planar distances between the planes parallel to the interface as the main process of relaxation. Moreover, for the $\Sigma 5(210)$, the mechanism was complemented by a rigid shift of one grain with respect to another in the [001] direction. Note that the shift was $16.8$\% of the lattice parameter.

The meaning of the optimization process is manifested in the reconstruction of the interface where individual positions of atoms are separated by a high energy barrier, and the efficiency of optimization method plays an important role in GB structure prediction. This is the case of the remaining two interfaces $\Sigma 17(410)$ and $\Sigma 13(510)$. The first feature which indicates the interface reconstruction is the comparison of LARGE GB energies obtained by SA and OPSA techniques (see table~\ref{tab2}). Application of an improved structure optimization technique OPSA leads to a significant decrease of GB energy. In particular, by $0.689$~J\,m$^{-2}$ for GB $\Sigma 17(410)$ and $0.831$~J\,m$^{-2}$ for GB $\Sigma 13(510)$.

\begin{figure}[!t]
\centerline{\includegraphics[width=0.65\textwidth]{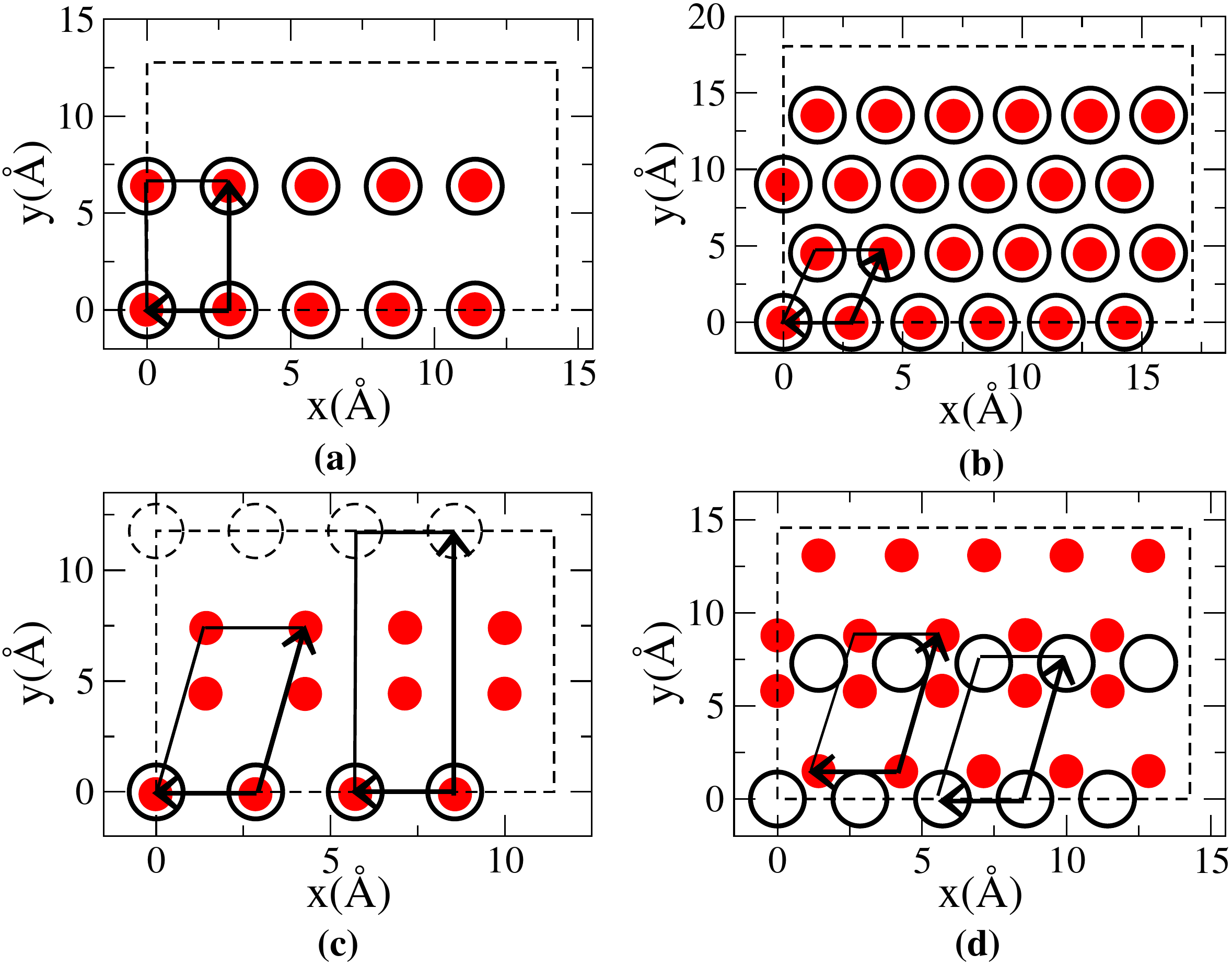}}
 \caption{(Color online) 2D-crystalline structure of GB planes (a) $\Sigma 5(210)$, (b) $\Sigma 5(310)$, (c) $\Sigma 17(410)$, (d) $\Sigma 13(510)$  before (black circles)
 and after (red dots) optimization. Corresponding 2D lattices are highlighted by arrows. } \label{GBplanes}
\end{figure}

The second feature is the changes in the crystalline structure of the interface plane after optimization. Figure~\ref{GBplanes} shows the position of atoms in the plane of GB before/after relaxation as a two dimensional (2D) crystalline structure of GB planes of all investigated GBs. It could be seen that 2D structure of GB planes $\Sigma 5(210)$ and $\Sigma 5(310)$ remained unchanged while the structure of  GB planes $\Sigma 17(410)$ and $\Sigma 13(510)$ has been significantly changed.  GB plane $\Sigma 5(210)$ is characterized by a rectangular Bravais lattice while $\Sigma 5(310)$ GB plane is characterized by oblique Bravais lattice. Note that both lattices contain one Fe atom in the basis. Lattice of $\Sigma 17(410)$ GB plane transfers from rectangular Bravais lattice with one Fe atom in the basis to oblique Bravais lattice with two Fe atoms in the basis. Lattice of $\Sigma 13(510)$ GB plane remained unchanged (oblique Bravais lattice). However, its basis changed from one Fe atom to two Fe atoms. This change could be explained by inward-relaxation of planes adjacent to GB plane, but a better performance of an improved optimization technique OPSA indicates much more complicated optimization process. It should be emphasised that for all the investigated GBs, the mirror symmetry is kept, which is not straightforward from figure~\ref{GBplanes}.

The third and decisive feature was discovered after a detailed analysis of the optimization process. An increased atomic density at GB planes $\Sigma 17(410)$ and $\Sigma 13(510)$ was accomplished by concurrent migration of individual atoms between planes $(410)$ resp. $(510)$ and $(001)$. Since this process cannot be described in terms of the changed inter-atomic distances or bond angles, it is defined as reconstruction. In case the reconstruction is not allowed by restriction of atomic migration in the direction [001], the resulting GB energies are $1.873$~J\,m$^{-2}$ and $1.934$~J\,m$^{-2}$ compared to those presented in table~\ref{tab2} for SMALL GBs $1.231$~J\,m$^{-2}$ and $1.070$~J\,m$^{-2}$. The dominant part of optimization with a restricted atomic migration in [001] direction is inter-granular shift along GB plane.

\begin{figure}[!b]
\centerline{\includegraphics[width=0.68\textwidth]{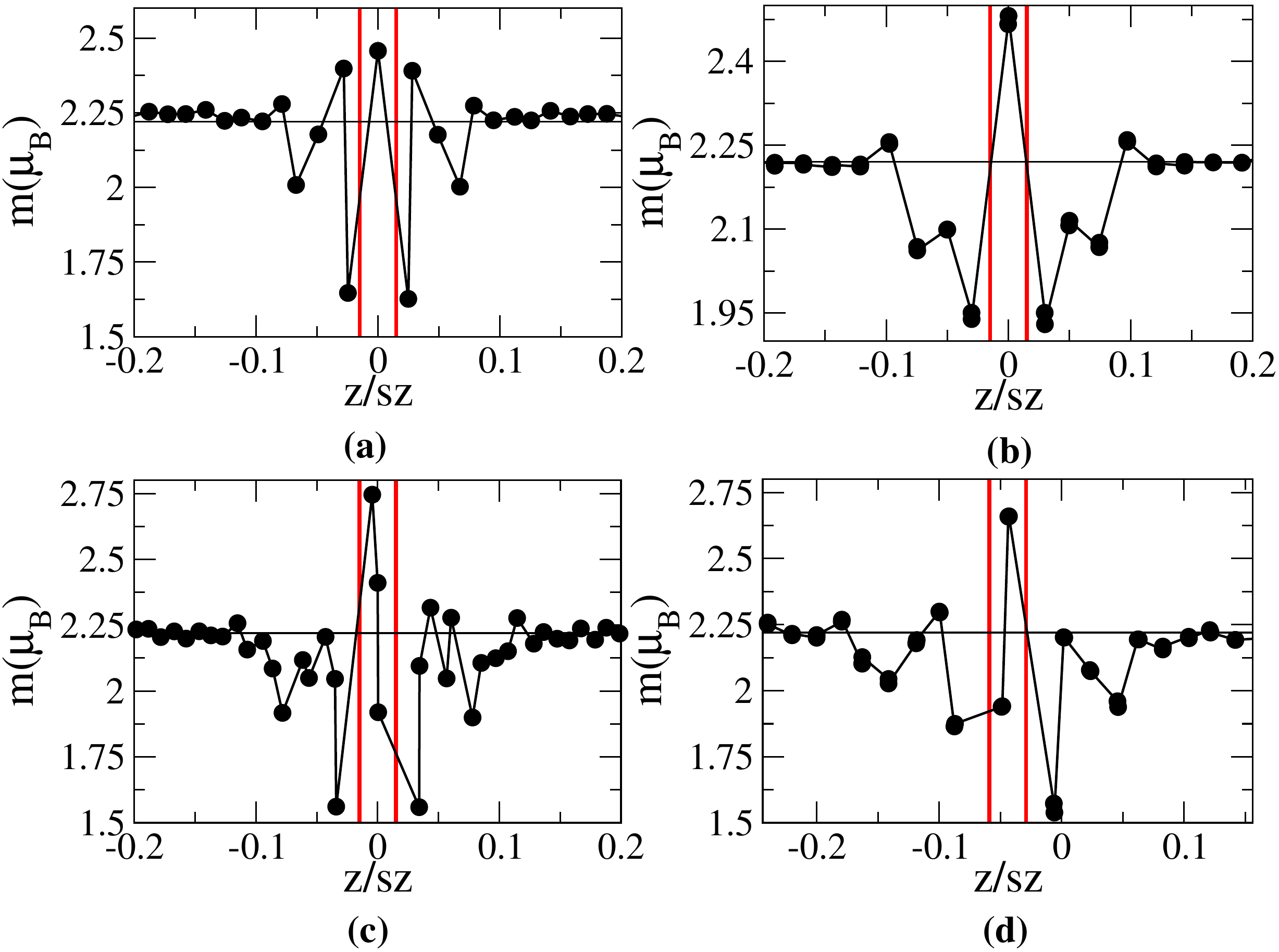}}
 \caption{(Color online) Variation of magnetic moment on Fe atoms in the direction perpendicular to the GB (a) $\Sigma 5(210)$, (b) $\Sigma 5(310)$, (c) $\Sigma 17(410)$, (d) $\Sigma 13(510)$. The positions of atoms are presented with respect to the supercell size ($sz$). Experimantal bulk magnetic moment value $2.22~\mu_\textrm{{B}}$ is indicated by horizontal solid line and GB plane is positioned between red vertical lines. } \label{mag}
\end{figure}

	The variation of magnetic moment of Fe atoms in the direction perpendicular to GB plane ($z$) for all the investigated GBs is shown in figure~\ref{mag}. The value of magnetic moment in the bulk area between interfaces agrees well with the experimental value of $2.22~\mu_\text{{B}}$. In the case of relaxed GBs $\Sigma 5(210)$ and $\Sigma 5(310)$, the magnetic moment on GB increases up to $2.46~\mu_\textrm{{B}}$ and $2.47~\mu_\textrm{{B}}$, respectively. The behaviour of local magnetic moment on reconstructed GBs $\Sigma 17(410)$ and $\Sigma 13(510)$ is not so straightforward. GB plane $(410)$ contains atoms with three different magnetic moment values 2.75, 2.41, 1.92~$\mu_\textrm{{B}}$ and GB plane $(510)$ atoms with two different magnetic moment values 2.66 and 1.94~$\mu_\textrm{{B}}$. To sum up, local magnetic moment is distributed uniformly on relaxed GBs and non-uniformly on reconstructed GBs.

\begin{figure}[!t]
\centerline{\includegraphics[width=0.40\textwidth]{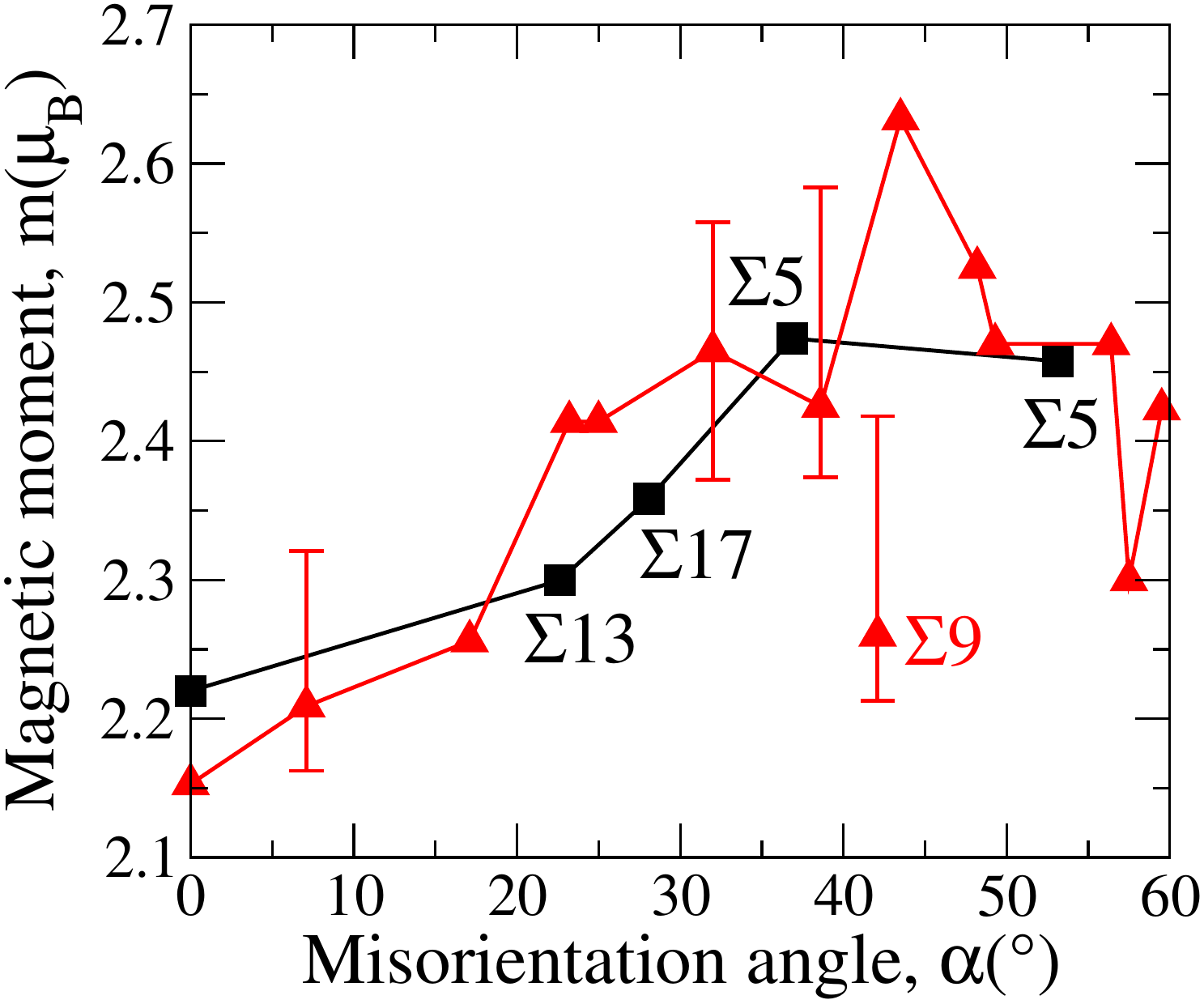}}
 \caption{(Color online) Simulated average local magnetic moment in GB plane as function of misorientation angle (squares) compared with the experimental result measured by Ii et al. \cite{ii13} (triangles). Line is just a guide for the eye. } \label{angle}
 \vspace{-2mm}
\end{figure}

In order to compare our data with experimental data published by Ii et al. \cite{ii13}, we computed the average local magnetic moment in the investigated GB planes. The local magnetic moment was averaged over all atoms positioned in GB plane. The dependence of the average local magnetic moment in GB plane as a function of misorientation angle for both simulated and experimental \cite{ii13} data, is in figure~\ref{angle}. Note that experimental data correspond to various types of GBs while in our case there were considered only $\langle100\rangle$ symmetric tilt GBs. The averaged local magnetic moment at GB plane increases up to $2.47~\mu_\textrm{{B}}$ and shows a maximum at misorientation angle $36.87^{\circ}$. The increasing tendency is in good agreement with experimentally observed data. We have demonstrated that the enhancement of the total magnetic moment on GB plane decreases not only as a result of the volume effect but also due to non-uniform distribution of local magnetic moment on GB plane. Non-uniform distribution of local magnetic moment was identified on low coincidence site density GBs $\Sigma 17(410)$ and $\Sigma 13(510)$ which undergo reconstruction. The same effect might be also responsible for low  magnetic moment enhancement measured at GB $\Sigma9$, which belongs to high $\Sigma$ GBs as GBs $\Sigma 17(410)$ and $\Sigma 13(510)$.
\vspace{-1mm}
\section{Conclusions}
\label{S:4}
It was shown that a prominent feature of certain GBs is the ability to reconstruct the structure during optimization process which in fact reduces the final energy. The effect of reconstruction was observed on low coincidence site density GBs  $ \Sigma 17(410)$ and $ \Sigma 13(510) $ and was accompanied by a significant reduction in energy. Reconstruction was allowed by migration of atoms from the plane $(001)$ to $(002)$. The migration of atoms consequently invokes an increase of atomic density in GB interface and the change of 2D crystalline structure at interface.

Response of local magnetic moment on the structure shows a uniform distribution in the relaxed GB planes and non-uniform distribution in reconstructed GB planes. The non-uniform distribution might be the reason for cusps on the magnetic moment --- misorientation angle curve measured by Ii et al. \cite{ii13}. The same phenomenon in addition to magneto-volume effect could be also the reason for a lower local magnetic moment in low-angle GBs.

We also proposed and applied an improved SA optimization technique, which in fact is a combination of parallel SA and GA algorithms. The main advantage of this technique is its ability to overcome the energy barriers that occur in the process of structure optimization. It was shown that GBs $ \Sigma 17(410)$ and $ \Sigma 13(510) $ are good candidates to test optimization algorithm qualities, because standard techniques like conventional SA algorithm are not capable of  overcoming the energy barriers related to the process of reconstruction. The second advantage is low dependence of the result of optimization process on initial adjustment of random number generator.
\section*{Acknowledgements}
This work was supported by Structural Funds of the European Union by means of the Research Agency of the Ministry of Education, Science, Research and Sport of the Slovak republic in the project ``CENTE I'' ITMS code 26240120011.

\newpage
\ukrainianpart
\title{Зернинномежова релаксація і реконструкція: вплив локального магнітного моменту}

\author{Е. Вітковска, П. Балло}
\address{Словацький технічний університет,   812 19 Братислава, Словаччина}

\makeukrtitle

\begin{abstract}
Ми представляємо детальне числове дослідження структури і локальних магнітних властивостей  симетричних похилених
зернинних меж $\langle 100 \rangle$ у  залізі з кубічною об'ємоцентричною структурою. Особлива увага приділяється зв'язку між типом
зернинномежової релаксації і локальними магнітними властивостями. Результати першопринципних обчислень показали, що зернинномежова
реконструкція приводить до неоднорідного розподілу локального магнітного моменту в зернинномежовій площині. Це протирічить результату,
отриманому в зернинномежовій площині, де спостережено просту релаксацію. Було досягнуто добре оптимізовані атомні  конфігурації поблизу
межі розділу за допомогою комбінації методу відпаленої оптимізації з генетичним алгоритмом.

\keywords залізо, зернинна межа, релаксація, реконструкція, магнітний момент,оптимізація
\end{abstract}

\begin{thebibliography}{10}
\providecommand{\url}[1]{\texttt{#1}}
\providecommand{\urlprefix}{URL }
\providecommand{\eprint}[2][]{\url{#2}}

\bibitem{mishin10}
Mishin Y., Asta M., Li J., Acta Mater., 2010, \textbf{58}, No. 4,
  1117; \bibdoi{10.1016/j.actamat.2009.10.049}.

\bibitem{siegel05}
Siegel D.J., Hamilton J., Acta Mater., 2005, \textbf{53}, 87; \bibdoi{10.1016/j.actamat.2004.09.006}.

\bibitem{wachowicz08}
Wachowicz E., Kiejna A., Comput. Mater. Sci., 2008, \textbf{43}, 736; \bibdoi{10.1016/j.commatsci.2008.01.063}.

\bibitem{ii13}
Ii S., Hirayama K., Matsunaga K., Fujii H., Tsurekawa S., Scr. Mater., 2013,
  \textbf{68}, 253; \\\bibdoi{10.1016/j.scriptamat.2012.10.028}.

\bibitem{hampel93}
Hampel K., Vvedensky D.D., Crampin S., Phys. Rev. B, 1993, \textbf{47}, No.~8,
  4810; \bibdoi{10.1103/PhysRevB.47.4810}.

\bibitem{wu96}
Wu R., Freeman A.J., Olson G.B., Phys. Rev. B, 1996, \textbf{53}, No.~11, 7504; \bibdoi{10.1103/PhysRevB.53.7504}.

\bibitem{cak08}
\v{C}\'{a}k M., \v{S}ob M., Hafner J., Phys. Rev. B, 2008, \textbf{78}, 054418; \bibdoi{10.1103/PhysRevB.78.054418}.

\bibitem{wachowicz10}
Wachowicz E., Ossowski T., Kiejna A., Phys. Rev. B, 2010, \textbf{81}, 094104; \bibdoi{10.1103/PhysRevB.81.094104}.

\bibitem{stoner38}
Stoner E.C., Proc. R. Soc. London, Ser. A, 1938, \textbf{165}, 372; \bibdoi{10.1098/rspa.1938.0066}.

\bibitem{manh09}
Nguyen-Manh D., Dudarev S.L., Phys. Rev. B, 2009, \textbf{80}, 104440; \bibdoi{10.1103/PhysRevB.80.104440}.

\bibitem{stekol02}
Stekolnikov A.A., Furthm\"{u}ller J., Bechstedt F., Phys. Rev. B, 2002,
  \textbf{65}, 115318; \bibdoi{10.1103/PhysRevB.65.115318}.

\bibitem{crljen03}
Crljen \v{Z}., Lazi\'{c} P., \v{S}ok\v{c}evi\'{c} D., Brako R., Phys. Rev. B,
  2003, \textbf{68}, 195411; \bibdoi{10.1103/PhysRevB.68.195411}.

\bibitem{ras12}
Abou-Ras D., Schaffer B., Schaffer M., Schmidt S.S., Caballero R., Unold T.,
  Phys. Rev. Lett., 2012, \textbf{108}, 075502; \bibdoi{10.1103/PhysRevLett.108.075502}.

\bibitem{persson03}
Persson C., Zunger A., Phys. Rev. Lett., 2003, \textbf{91}, No.~26, 266401; \bibdoi{10.1103/PhysRevLett.91.266401}.

\bibitem{yang15}
Yang L., Gao F., Kurtz R.J., Zu X.T., Acta Mater., 2015, \textbf{82}, 275; \bibdoi{10.1016/j.actamat.2014.09.015}.

\bibitem{morris96}
Morris J.R., Fu C.L., Ho K.M., Phys. Rev. B, 1996, \textbf{54}, No.~1, 132; \bibdoi{10.1103/PhysRevB.54.132}.

\bibitem{du11}
Du Y.A., Ismer L., Rogal J., Hickel T., Neugebauer J., Drautz R., Phys. Rev. B,
  2011, \textbf{84}, 144121; \\\bibdoi{10.1103/PhysRevB.84.144121}.

\bibitem{zhang09}
Zhang J., Wang C.Z., Ho K.M., Phys. Rev. B, 2009, \textbf{80}, 174102; \bibdoi{10.1103/PhysRevB.80.174102}.

\bibitem{dugan09}
Dugan N., Erko\c{c} \c{S}., Comput. Mater. Sci., 2009, \textbf{45}, 127; \bibdoi{10.1016/j.commatsci.2008.03.045}.

\bibitem{amy15}
Kaczmarowski A., Yang S., Szlufarska I., Morgan D., Comput. Mater. Sci., 2015,
  \textbf{98}, 234;\\ \bibdoi{10.1016/j.commatsci.2014.10.062}.

\bibitem{kirk83}
Kirkpatrick S., Gelatt C.D., Vecchi M.P., Science, 1983, \textbf{220}, No.
  4598, 671; \bibdoi{10.1126/science.220.4598.671}.

\bibitem{daw84}
Daw M.S., Baskes M.I., Phys. Rev. B, 1984, \textbf{29}, No.~12, 6443; \bibdoi{10.1103/PhysRevB.29.6443}.

\bibitem{mendelev03}
Mendelev M.I., Han S., Srolovitz D.J., Ackland G.J., Sun D., Asta M., Philos.
  Mag., 2003, \textbf{83}, 3977;\\ \bibdoi{10.1080/14786430310001613264}.

\bibitem{gonze09}
Gonze X., Amadon B., Anglade P.-M., Beuken J.-M., Bottin F., Boulanger P., Bruneval F., Caliste D., Caracas~R., C\^{o}t\'{e}~M., et~al., Comput. Phys. Commun., 2009, \textbf{180}, 2582; \bibdoi{10.1016/j.cpc.2009.07.007}.

\bibitem{troullier91}
Troullier N., Martins J.L., Phys. Rev. B, 1991, \textbf{43}, No.~3, 1993; \bibdoi{10.1103/PhysRevB.43.1993}.

\bibitem{goedecker96}
Goedecker S., Teter M., Hutter J., Phys. Rev. B, 1996, \textbf{54}, No.~3,
  1703; \bibdoi{10.1103/PhysRevB.54.1703}.

\bibitem{kittel96}
Kittel C., Introduction to Solid State Physics, Wiley-Interscience, New York,
  1996.

\bibitem{pearson67}
Pearson W.B., A Handbook of Lattice Spacings and Structures of Metals and
  Alloys, Pergamon Press, Oxford, 1967.

\end{thebibliography}
\end{document}